# North Sea Wind Power Hub:
## System Configurations, Grid Implementation and Techno-economic Assessment

G. MISYRIS[1]*, T. VAN CUTSEM[2], J. G. MØLLER[1], M. DIJOKAS[1], O. RENOM ESTRAGUÉS[1],
B. BASTIN[2], S. CHATZIVASILEIADIS[1], A. H. NIELSEN[1], T. WECKESSER[3], J. ØSTERGAARD[1],
F. KRYEZI[4]

[1]Technical University of Denmark (DTU), [2]University of Liège, [3]Dansk Energi, [4]Energinet
Denmark, Belgium

**SUMMARY**

In 2017, Energinet and TenneT, the Danish and Dutch Transmission System Operators (TSOs), have announced the North Sea Wind Power Hub (NSWPH) project. The project aims at increasing by 36 GW the North Sea offshore wind capacity, with an artificial island collecting all the power produced by wind turbines and several HVDC links transmitting this power to the onshore grids. This project brings together new opportunities and new challenges, both from a technical and economic point of view. In this regard, this paper presents three analyses regarding the design and operation of such an offshore system.

First, we perform a techno-economic assessment of different grid configurations for the collection of the power produced by wind farms and its transmission to the hub. In this analysis, two frequencies and two voltage levels for the operation of the offshore grid are investigated. Our findings show that the nominal-frequency high-voltage option is the more suitable, as low-frequency does not bring any advantage and low-voltage would results in higher costs.

The second analysis is related to the differences in operating the system with low- or zero-inertia; different dynamic studies are performed for each configuration to identify proper control actions and their stability properties. Comparing the outcomes of the simulations, we observed that voltage and frequency oscillations are better damped in the zero-inertia system; however, the risk of propagating offshore faults in the connected onshore grids is mitigated with the inclusion of the synchronous condensers.

Lastly, a comparison of ElectroMagnetic Transient (EMT) and phasor-mode (also known as RMS) models is presented, in order to understand their appropriateness of simulating low- and zero-inertia systems. The results show that phasor approximation modelling can be used, as long as eigen-frequencies in power network are well damped.

*gmisy@elektro.dtu.dk

# I. INTRODUCTION

National and international ambitions of reducing greenhouse gas emissions result in Renewable Energy Sources (RES) progressively replacing conventional generation based on fossil fuels [1]. In 2017, representatives of the Dutch, Danish and German Transmission System Operators (TSOs) established a consortium to explore the possibility of developing a North Sea Wind Power Hub (NSWPH). This project aims at increasing the North Sea offshore wind capacity by integrating additional 36 GW of wind power, with an artificial island collecting all the power produced by wind turbines and several High Voltage Direct Current (HVDC) links transmitting this power to Denmark, the Netherlands, Norway, Germany and the UK [2].

Due to the length of the submarine power cables, the HVDC technology is the only solution to build a cost-effective transmission infrastructure. Moreover, Voltage-Source Converters (VSC) allow for high flexibility and improved control capabilities. Thus, the connections between the offshore island and the onshore grids will be formed by multiple point-to-point VSC-HVDC links; although considered at the beginning, the option of a multi-terminal HVDC grid [3] was discarded due to the lack of experience with HVDC breakers. The function of a NSWPH would be twofold: (i) increase the shares of renewable generation in the involved countries, and (ii) allow for power exchanges between partner TSOs by means of additional transmission capacity.

Such a project brings together a series of new opportunities and new challenges, both from a technical and economic point of view. Starting from the economic aspects, the main question is whether the NSWPH shall operate at nominal AC frequency (50 Hz) or at Lower Frequency AC (LFAC), e.g. 16.67 Hz. Considering that the maximum net power of NSWPH could go up to 36 GW, several medium- and high-voltage AC cables will be deployed for the connection of wind turbines to the hub. Hence, reactive power compensation is necessary to improve power quality and voltage profiles in proximity of the island. At lower frequency, series reactances and shunt susceptances of power cables are lower compared to their values at nominal frequency (50 Hz). On the one hand, the distance between the island and the wind farms can be extended without the need of additional compensation. On the other hand, transformers and inductors increase in size when designed to operate at lower frequencies, which imposes an additional constraint in terms of footprint on the artificial island.

The technical aspect of this study concerns the growing use of non-synchronous converter-based resources. Integrating these resources affects the reliability of the power system and changes the nature of grid dynamics and control. Indeed, the replacement of synchronous generators with inverter-based resources is followed by a decrease of system inertia. When the replacement is only partial, the system is referred to as low-inertia power system. In case of complete absence of synchronous generation, it becomes a zero-inertia system. This poses the challenge of faster active power sharing between the converters of the HVDC links.

Another challenge concerns voltage control in the proximity of those converters. Indeed, the control systems of Power Electronic (PE) devices usually rely on measurements of voltage magnitudes and phase angles at their terminals. In high-inertia systems, voltage and frequency are stiff, therefore voltage magnitudes and angles at the terminals are not largely affected. Onshore grids usually host multiple synchronous generating units, that inherently provide a stiff voltage and frequency [4]; this does not happen in offshore grids, which usually consist only of multiple PE devices. Particularly, active power will be generated by (Type IV) wind turbines, then gathered through submarine cables in various offshore converter platforms and lastly transferred through HVDC links to the interconnected onshore grids [5]. Consequently, in such zero-inertia offshore systems, voltage magnitudes and phase angles are relatively sensitive to current injections from the wind turbines.

Furthermore, harmonic instability challenges [6] arise when connecting PE resources to grids without synchronous generators because of the interaction between converter control actuation and eigen-frequencies in power networks. Converters are commonly equipped with multiple time-scale control systems for regulating current flows at their terminal and power exchanges with the grid. To analyse the stability of a zero-inertia system and consider harmonic instability phenomena, both voltage and signal processing dynamics need to be included in the model. On the one hand, detailed modelling of the system dynamics increases the reliability of stability analyses; on the other hand, it increases the computational time required to solve the differential algebraic system of equations describing grid dynamics. Large computing times make security assessment more difficult, e.g. it takes more time to perform an analysis of post-disturbance system conditions in order to verify that power ratings and voltage constraints are violated.

To decrease computing times, system operators could resort to phasor-mode simulation, widely used in AC systems without large PE devices. In the latter, network lines, converters' inner controllers and signal processing dynamics are neglected. On the one hand, by neglecting these dynamics, models under phasor approximation are unable to capture harmonic instability phenomena; on the other hand, they are suitable for capturing slower phenomena associated with the dynamics of converters' outer controllers (as far as there is a distinct time separation between converters' outer controllers and voltage dynamics). The need for fast, accurate and representative models for performing time-domain simulation analysis in grids with high penetration of PE devices calls for the evaluation under which conditions the phasor approximation is still able to predict the system response after a disturbance. Identifying those conditions will allow performing online security assessment with faster phasor-mode simulation tools for a (hopefully large) subset of disturbances, while keeping the computationally demanding EMT for the other, more severe disturbances.

The aim of this paper is to provide the groundwork for understanding the non-linear behaviour of offshore AC networks, as well as novel insights in their fundamental stability and control properties. More specifically:

- We perform a techno-economic analysis to identify the economic benefit of offshore grids operated at low and normal frequency; this includes both the costs for the components as well as the maximum power transfer capability of the cables.
- We investigate and compare a zero- (100% inverter-based grid) and a low-inertia (synchronous condensers on the island) configuration, where the stability properties of each system and their responses to large disturbances are investigated.
- We assess the limitations of simulation software techniques for low- and zero- inertia systems, to examine when EMT simulations become crucial in capturing phenomena that the phasor-mode approximation cannot.

The rest of the paper is organized as follows. Section 2 presents the concept of NSWPH and the study case considered in this paper. In Section 3, first, a techno-economic analysis is carried out to identify the preferable frequency and voltage level for the offshore AC grid. Then, two different configurations of the offshore system are presented. Section 4 presents a comparison between the two considered configurations, with reference to their respective stability properties. Section 5 presents a comparison of EMT and phasor-mode (also known as RMS) models. Conclusions are drawn in Section 6.

## II. NORTH SEA WIND POWER HUB (NSWPH): GRID TOPOLOGY

The NSWPH will follow a modular Hub-and-Spoke concept, according to NSWPH consortium, which is the common practice for large-scale offshore wind deployment. This concept relies on the progressive construction of modular hubs in the North Sea which are connected to the countries in Northern Europe by means of HVDC lines.

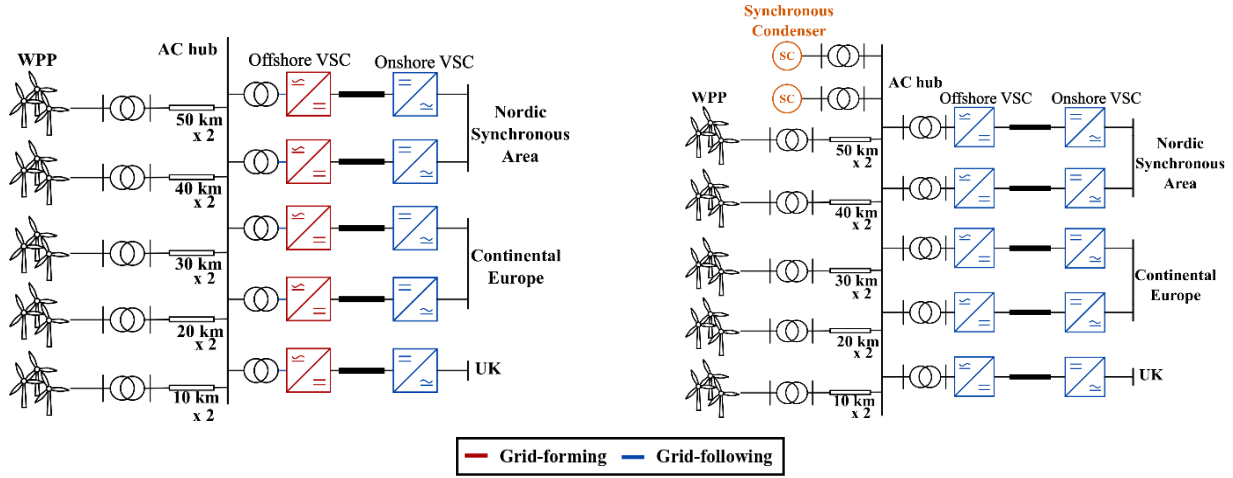

*Figure 1: Zero-inertia (left) and low-inertia (right) topologies for NSWPH.*

The grid layout used for the analyses of this paper is depicted in Figure 1. This grid configuration is based on the topology that the TSOs involved in the project are considering at the moment. In this layout, the wind power generated by the wind farms is transferred through High-Voltage-Alternating-Current (HVAC) cables (220 kV, with 400 MVA nominal power) to the hub. The cable parameters are taken from manufacturer data sheets [7]. Five HVDC links are used to transfer onshore the amount of power. The onshore grid is represented by a grid equivalent with inertial and primary response [8], [9]. The converters on the offshore side operate in grid-forming (zero-inertia) or grid-following (low-inertia) mode. As base power, we consider $S_b$ = 1000 MVA. We assume the rated power of offshore converters is 1100 MVA, while the rated power of wind farms is 700 MVA. In the low-inertia configuration, the rated power of each of the two synchronous condensers is 350 MVA. This model represents a small system compared to the maximum total power capacity envisioned for the NSWPH, but it equally captures all relevant dynamic phenomena and provides a good overview of the instability mechanisms.

### III. TECHNO-ECONOMIC ANALYSIS OF OFFSHORE AC COLLECTION GRID OPTIONS

One of the major technical limitations of long-distance AC submarine transmission is the charging current, caused by the shunt capacitance of the cables, which results in a huge amount of reactive power produced. For this reason, conventional 50/60 Hz AC transmission cables connecting remote offshore wind farms to onshore grids require reactive compensation in such proportions that HVDC is often the preferred option [10].

As an alternative, authors in [11], [12], [13] have proposed to operate wind farms at lower frequencies to reduce effects of line charging and permit longer transmission distances. However, the design of grid components depends on the operating frequency of the system. In this chapter, we present a techno-economic assessment of operating collection grids and transmission systems at different frequencies and voltage levels.

### III.i. Design Considerations

For the techno-economic assessment of the collection grids and transmission cables, two voltage levels (66 and 220 kV) and two system frequencies (16.67 and 50 Hz) are considered. In order to evaluate the different options, a selection of power cables and power transformers has been made.

The power cables are three-core cables with copper conductors: to make calculation easier, each of them is assumed to be a solid cylinder with cross-section of 630 mm$^2$ and diameter of 28.3 mm. The nominal current of the cables is 1.05 kA, while the power rating varies with the voltage level, respectively 120 and 400 MVA (for 66 and 220 kV). The model proposed in [14] is adopted for assessing the cost of the cables, which are shown in Table 4. Cable installation costs have been chosen equal to 0.345 Mill. Euros per kilometer as suggested in [15].

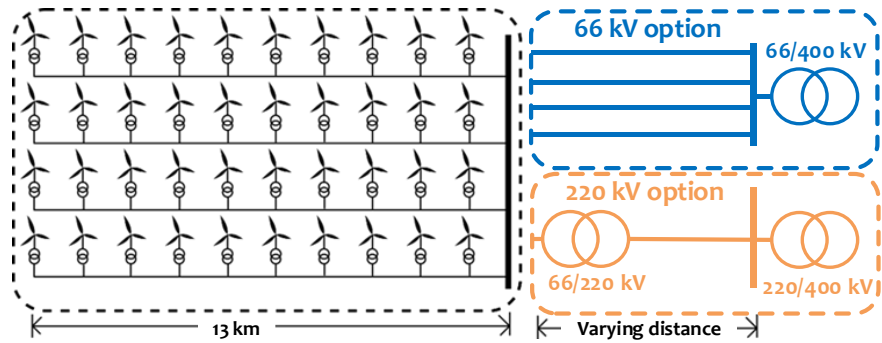

*Figure 2: Wind farm layout and options of transmission voltage*

The transformers are core-type transformers with different power and voltage ratings. For the 66-kV option, each turbine is connected to the transmission system through a 0.67/66-kV transformer, which has a power rating of 10 MVA. A 66/400-kV transformer is then located at the hub, in the proximity of the HVDC converter station; its power rating is 400 MVA. For the 220-kV option, the voltage is stepped up by a 66/220-kV transformer with power rating of 400 MVA and then again in proximity of the hub, with a 220/400-kV 400-MVA transformer. The two configurations are depicted in Figure 2. In the following part, the pros and cons of each grid layout are discussed.

As mentioned above, the transmission of active power through long-distance AC cables is limited by the reactive power produced by their shunt capacitance. Figure 3 shows the maximum power transfer for the four transmission options assuming a charging capacitance of 0.2 µF/km. The reactive power produced by shunt capacitances is proportional to the system frequency [16]. Operating a transmission cable at one third of the frequency would lead to a one-third reduction of the shunt reactance. However, the shunt inductance is the main component of the cost of reactive compensation [17]. At 16.67 Hz, the inductance would be exactly equal to the inductance at 50 Hz. This relation is valid for short distances, where reactive compensation is performed at the two terminals of the cable. For transmission distances far beyond 50 km, 50-Hz operation would require compensation equally distributed along the length of the cable. This would significantly increase the cost of the 50-Hz alternative.

Transmission losses are also impacted by the frequency, as the skin-effect of the current increases at higher frequencies. The skin-depth, that is the depth at which the current density drops significantly, is proportional to the reciprocal of the square root of the frequency [18]. The smaller is the skin-dept of a cylindrical conductor, the greater is its AC resistance [18]. Thus, lowering the operating frequency will reduce joule losses in transmission cables. Annual losses and the resulting costs for the two frequency options are shown in Table 2. For the calculation of annual losses, a utilization factor of 0.5 is assumed, resulting in 4380 full-load hours per year. For the cost of losses, the price of electricity is assumed to be 30 Euro/MWh, similar to the five-year average of Denmark. An estimation of the total losses can be obtained by multiplying the average distance between the wind farms and the hub by the losses per km at 50 or 16.67-Hz.

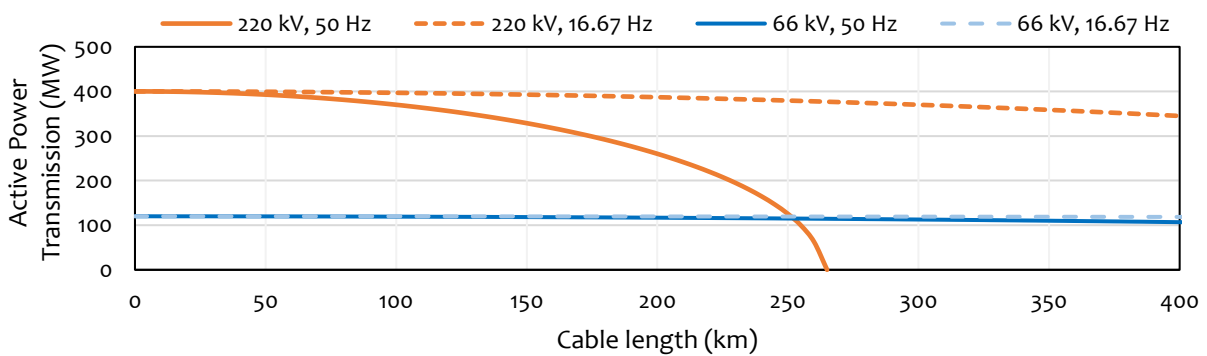

*Figure 3: Maximum active power transfer depending on the length of the transmission cable.*

It must be said that, in order to reduce the skin-effect, it is common practice to use stranded wires instead of solid conductors. In this analysis, we consider solid conductors to make the calculations easier; the presented costs are most probably an overestimation of the actual costs but represent a plausible basis for comparison of the two frequency options.

Concerning power transformers, their cost and size depend on the operating frequency. Indeed, the core cross-sectional area of a transformer is inversely proportional to the system frequency [19]. It follows that, by reducing the system frequency to 16.67-Hz, the core cross-section increases threefold. Moreover, also the length of the windings increases because of the greater circumference of transformer legs. The cost and the required mass of active materials for the different transformers are displayed in Table 1. In the table, $B_{max}$ is the saturation flux density of the core material, $J_{max}$ is the maximum winding current density, used to compute winding cross-sections, and $dV_{max}$ is the maximum voltage drop per turn, used for computing the minimum number of turns. The price of copper and steel is assumed to be respectively 7 and 3 Euros per kilogram. Additional cost of casing, civil engineering, transport and installation must be assumed to scale in proportion with the mass.

The cost of a 500-MW 220-kV AC platform is provided in [15]. This is used as basis for estimation of the cost of the 400-MW platform to be used in the 50-Hz 220-kV alternative. For the 16.67-Hz option, the cost has been scaled by a factor of 3, as it is assumed to be proportional to the load of the platform. A detailed structural design would be required to obtain the true cost for the low-frequency option; however, this is out of the scope of this study. The costs of offshore AC platform are shown in Table 3. On the one hand, 66-kV transmission could avoid certain costs, e.g. for the offshore transformer platforms, as it is expected that wind turbines will be able to operate at this voltage in near future [10]. On the other hand, increasing the voltage to 220-kV would significantly reduce transmission losses and the number of cables required to collect the power produced by wind turbines.

| Voltage level (kV) | Power rating (MVA) | Cost (mill. Euros / km) |
|---|---|---|
| 66 | 120 | 0.72 |
| 220 | 400 | 1.31 |

Table 4: Cost of power cables.

|  | 16.67 Hz | 50 Hz |
|---|---|---|
| Jacket | 27,6 | 9,2 |
| Topside | 66 | 22 |
| Installation | 22,08 | 7,36 |
| **Total** | **115,68** | **38,56** |

Table 3: Cost of offshore AC platform (mill. Euros).

|  | DC | 16.67Hz | 50.0Hz |
|---|---|---|---|
| Resistance (mΩ/km) | 29.5 | 29.7 | 34.9 |
| Full load losses, $3I_n^2R$ (W/m) | 97.6 | 98.2 | 115.4 |
| **Annual losses (MWh/km)** | **427** | **430** | **506** |
| **Cost of 20-year losses (mill. Euro/km)** | **0.256** | **0.258** | **0.303** |

Table 2: Cable resistances, annual losses and related costs.

| Transformer design parameters ||||| Mass (T) ||| Cost (Mill. Euro) |||
|---|---|---|---|---|---|---|---|---|---|---|
| Voltage rating (kV) | Power rating (MVA) | $B_{max}$ (T) | $J_{max}$ (A/mm²) | $dV_{max}$ (V) | 16.67 Hz | 50.0 Hz | Diff. | 16.67 Hz | 50.0 Hz | Diff. |
| 0.67/66 | 10 | 1 | 7 | 22 | 27 | 9 | **18** | 0,10 | 0,04 | **0,06** |
| 66/220 | 400 | 1 | 3 | 220 | 1075 | 371 | **704** | 3,75 | 1,49 | **2,27** |
| 220/400 | 400 | 1 | 3 | 220 | 1387 | 489 | **897** | 4,78 | 1,93 | **2,85** |
| 66/400 | 400 | 1 | 3 | 220 | 1207 | 417 | **791** | 4,16 | 1,64 | **2,52** |

Table 1: Mass and cost of active material (core and windings) for a range of medium and high voltage transformers.

**III.ii. Economic Assessment**

A cost estimation model is developed for comparing the Total Cost of Ownership (TCO) of the different option for a time window of 20 years. The capital costs included in model comprise the cost of active materials of transformers and their platforms (if present), the cost of the cables and their installation and the 20-year amortization of all investments at 2% interest. Cost components which are invariant of design frequency and transmission voltage, such as cost of array cables and turbines, are excluded from the study. Moreover, the cost of losses is included in the model, considering losses in the transmission cables and in the collection grids. The losses are calculated for a time period of 20 years. The TCO of the four grid options are shown in Figure 4.

For the NSWPH project, the power collected by the hub will probably be in the range of 10-15 GW [20], although alternatives up to 36 GW are also under consideration [21]. The amount of power collected by the hub has an impact on the length of power cables. Indeed, the area covered by the turbines, and thus the maximum transmission distance between these and the hub, increases with the amount of power produced. From Figure 4, it is evident that the transmission distance plays an important role on the decision of the type of transmission technology.

The test system used in the remainder of this paper is a 4 GW hub surrounded by 5 wind farms (WF) with 800 MW of installed capacity. Assuming a maximum wind power production of 6W/m2 [22], the resulting area covered by the wind farms is, at least, 667 km2. This corresponds to a circular area with a radius of 14.5 km. If the same calculation is done for a hub collecting 36 GW of wind power, the corresponding radius would be in the range of 40-50 km. These distances seem far too small for low frequency AC transmission to be a relevant option for the offshore grid. Below 50 km, the line charging of both the 220kV and the 66kV cables is simply too small to justify a change of standards.

Figure 4 compares the magnitude of those cost components which depend on either design frequency or transmission system voltage. It can be noted that for the 66-kV option, there is a very small difference between TCO of the low- and normal frequency alternatives. At short transmission distances, the 50-Hz alternative is more convenient due to smaller investments in transformers, while the 16.67-Hz alternative is more convenient for longer transmission distances (above 30km) due to smaller losses. It is worth mentioning that for the 66-kV option and short transmission distances, the 16.67-Hz alternative is slightly more convenient than the 220-kV option and 50-Hz alternative. However, normal design frequency and 220-kV transmission is the preferred option for transmission distances of about 15 km due to the excessive cost of cables in the 66-kV option. The combination of high voltage and low frequency transmission is not a cost-effective alternative for transmission distances below 60 km.

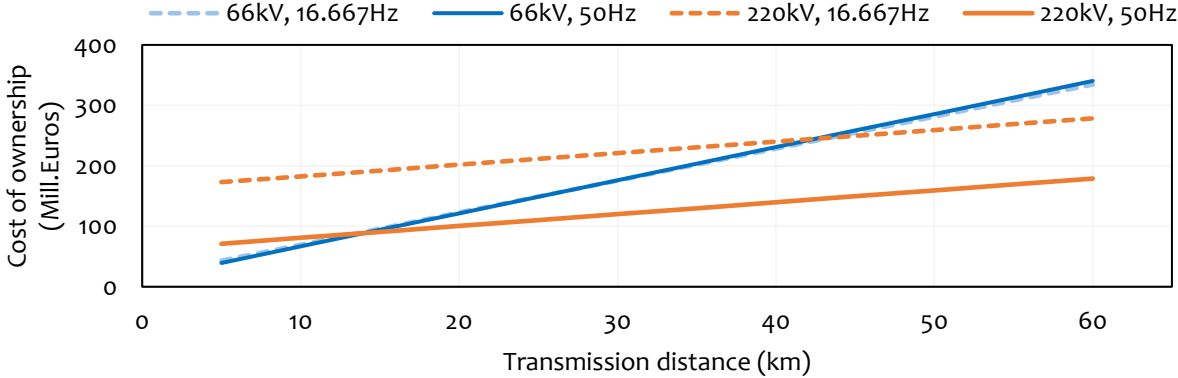

*Figure 4: Comparison of cost of ownership between transmission system options for a 400 MW wind farm.*

Based on the findings of this economic analysis, in the following section it is assumed that the AC offshore grid is operated at 50 Hz and 220 kV.

## IV. TECHNICAL ANALYSIS OF OFFSHORE AC GRID CONFIGURATION

In case of an AC grid on the island and between the wind farms, there are two possible configurations, namely a zero- and a low-inertia solution. The zero-inertia solution corresponds to a 100% converter-based system (see Figure 1 - left) while the low-inertia solution corresponds to a system dominated by PE devices, but with at least one synchronous condenser connected (see Figure 1 - right). The purpose of the latter is to provide: (i) a voltage source to which the VSCs synchronise in grid following mode, and (ii) a buffer of kinetic energy, stored in the rotating masses (rotor, possibly complemented by a flywheel). The speed of rotation of the synchronous condenser sets the frequency of the offshore AC grid. On the onshore side, each converter operates in grid-following mode [3], where it provides constant reactive power and regulates the voltage of the HVDC-link. A generic wind farm model is used, based on the dynamic equivalent presented in [23]. No special control is required from the wind generators (grid-following control), which operate in maximum power tracking mode, and under unity power factor.

The main difference between the two configurations lies in the control of the offshore VSCs and the components placed on the island (see Figure 5 left and Figure 5 right, respectively).

### IV.i. Zero-inertia configuration

The zero-inertia configuration includes multiple grid-forming converters [24] operating in parallel, controlling the voltage at their terminals. Figure 5 (left) depicts the basic control structure of a grid-forming converter, which consists of the active power controller and the virtual impedance controller. Their main principles are to impose the frequency and the AC voltage on the offshore AC grid.

Multiple grid-forming VSCs operating in parallel be equipped each with a frequency droop control scheme [5], which is automatically activated to adjust the active power injected or absorbed by the VSCs after a disturbance. The objective is to distribute the change in power among the converters in the zero-inertia systems based on pre-defined participations. In that process, the system frequency is used as a communication signal between the converters. We can infer that the operating principle of a grid-forming converters is similar to the one of synchronous generators, apart from the hard-current limits of the converters. However, due to the lack of an energy buffer, a sudden power imbalance requires an almost immediate reaction of the VSCs. As a result, fast control actions are necessary to stabilize the system, considering though the current limits of the converters.

Moreover, a voltage set-point is used to control the output voltage of the offshore converter. The reactive power injected/absorbed by the offshore converter reacts to voltage deviations on the Point of Common Coupling (PCC) caused by reactive power imbalances in the system. Lastly, the virtual impedance controller is also employed for damping the grid-frequency resonant poles and electromagnetic transients caused by the multiple cables of the offshore grid [25].

### IV.ii. Low-inertia configuration

The low-inertia configuration combines the advantages of conventional generation and VSC technology [26]. It consists of synchronous condensers connected to the offshore grid and multiple grid-following converters [26]. The synchronous condensers set the frequency and the voltage at the PCC. Any mismatch in active power on the island is reflected on a change of rotor speed of the synchronous condensers. The grid-following converters closely track the frequency set by the synchronous condensers, collect the wind power and transfer it through the HVDC links to the onshore grids. As aforementioned, the synchronous condensers act as an energy buffer in the case of a power imbalance. This buys some time to re-dispatch power to the grid-following converters. Dynamic simulations have shown that the system can stand the outage of one of the two synchronous condensers, i.e. it operates in a stable manner with a single 350-MVA machine [26].

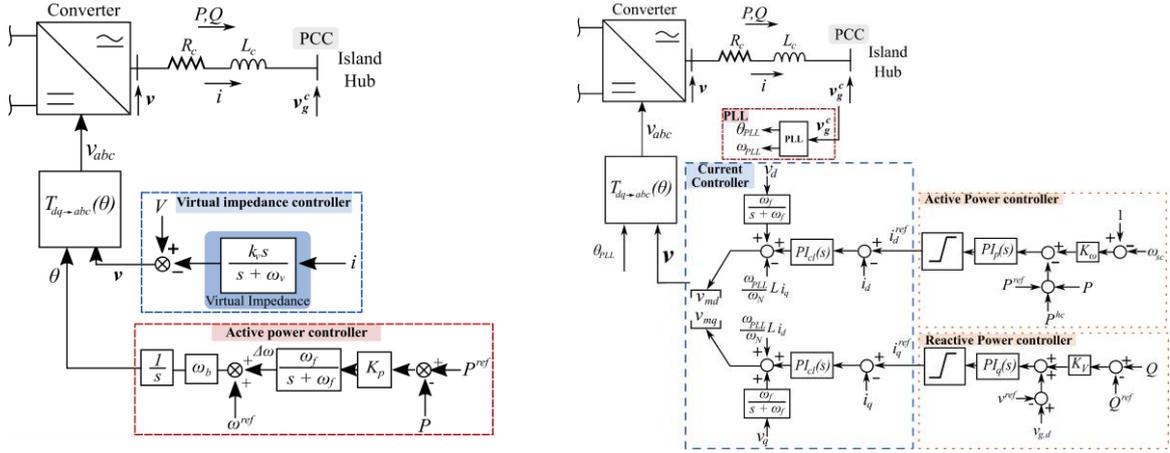

*Figure 5: Control principles of grid forming offshore converters for the zero-inertia case (left) and grid-following converter for the low-inertia case (right).*

To restore the active power balance and the stored kinetic energy of the synchronous condenser, the offshore converters must be able to adjust the power extracted from the offshore grid. This adjustment has to be performed with specific time response; not too slow, in order to avoid excessively large frequency deviations but not too fast either, in order to avoid large DC-voltage deviations of the HVDC-link, as well as fast disturbance propagations to the onshore grids. Thus, additional frequency droops (see Figure 5 - right) are added to their outer controllers to enable their participation to frequency regulation of the grid. A reactive power-voltage droop control is used for the participation of the VSC converters in the hub voltage control (also controlled by the AVRs of the synchronous condensers). The resulting offshore converter control is depicted in Figure 5 (right) and is similar to the one presented in [26]. The main components of the control unit are the PLL, the outer active and reactive controllers and the inner controllers.

Additionally, a centralized frequency controller for active power regulation is considered to return to nominal frequency after a disturbance [26]. Its principle is to provide an active power correction for each of the offshore VSC based on the deviation of the synchronous condenser's rotor speed. It should be noted that no time-delay is considered on the frequency (rotor speed) measurement, since the distance between the offshore converters and the synchronous condenser is very short, and the centralized controller is rather slow. Another advantage of the centralized frequency controller is the additional degree of freedom for sharing the active power among the converters (in the zero-inertia case the active power sharing depends only on the frequency droop values of the converters).

### IV.iii. Comparison between low- and zero- inertia configurations

In this section, a comparison between the two configurations is performed, in order to show their stability properties and illustrate their ability to restore the voltage and the frequency of the offshore system for different type of disturbances. For comparison purposes, we define the following criteria for secure operation of the offshore system:

1. Maintain nearly constant voltage at the AC hub,

2. Regulate the frequency of the offshore system,
3. Avoid fast disturbance propagation to the onshore grids.

Moreover, we evaluate them based on three different scenarios:

1. A 200 MW power request from one of the onshore grids interconnected to the island, where active power is transferred from the rest of the interconnected onshore grids. The active power sharing depends on the frequency droops. For our studies, the offshore converters are tuned homogeneously. Hence, the frequency droops have the same values, and the requested active power is distributed equally among the converters.
2. Offshore HVDC converter outage, i.e. HVDC converter trip, where one of the offshore converters is disconnected without considering a DC fault. With this scenario, we validate how fast a disturbance is propagated to the interconnected onshore grids. Considering that future onshore grids will operate with lower inertia, fast disturbance propagation may cause large frequency deviations.
3. Wind power loss corresponding to the disconnection of the furthest wind farm. With this scenario we validate the ability of both configurations to stabilize and restore the Hub-voltage and how fast the disturbance propagates to the onshore grids.

The performance of the zero- and low-inertia configuration are assessed based on EMT simulations for all 3 scenarios.

Figure 6 presents voltage evolution at the AC-Hub bus in the aforementioned scenarios. As it can be seen, both configurations are able to maintain voltage after all three disturbances. It can also be noted that the disconnection of the furthest wind farm causes the largest voltage deviation compared to the other scenarios. This was expected since both cables were fully loaded at pre-fault state, causing a large change of the reactive power at the AC-Hub bus. Moreover, the voltage presents an oscillatory behavior. The frequency of these oscillations depends on the cable parameters and the damping ratio of these oscillations depends on the tuning of the virtual impedance controller. Lastly, for the considered scenarios, we can conclude that in the zero-inertia configuration the voltage responses are better damped and a new steady state is reached faster than in the low-inertia configuration.

In the zero-inertia configuration, due to the absence of energy storage, a disturbance propagates almost instantaneously to the interconnected onshore grids. Considering scenario 2 and the system response depicted in Figure 7 (middle plot), where an offshore converter outage occurs, the active powers absorbed by the remaining offshore converters change to a new steady state (post-fault equilibrium) within a few milliseconds. In the low inertia case, due to the kinetic energy stored in the rotating mass of the synchronous condenser, the disturbance in scenario 2 and 3 propagate slower to the interconnected onshore grids. As can be seen in Figure 7, the rate of change of active power delivered to the interconnected onshore grids is much lower compared to the zero-inertia case. Thus, we can infer that large disturbances taking place in the offshore grid have less severe impact on the frequencies of the interconnected onshore grids. Lastly, considering scenario 1 and the results depicted in Figure 7 (left plot), we can infer that both configurations can fast provide active power to a connected onshore grid. In the zero-inertia system a fast response with short settling time can be observed, while in the low-inertia configuration the power request resulted in an overshoot.

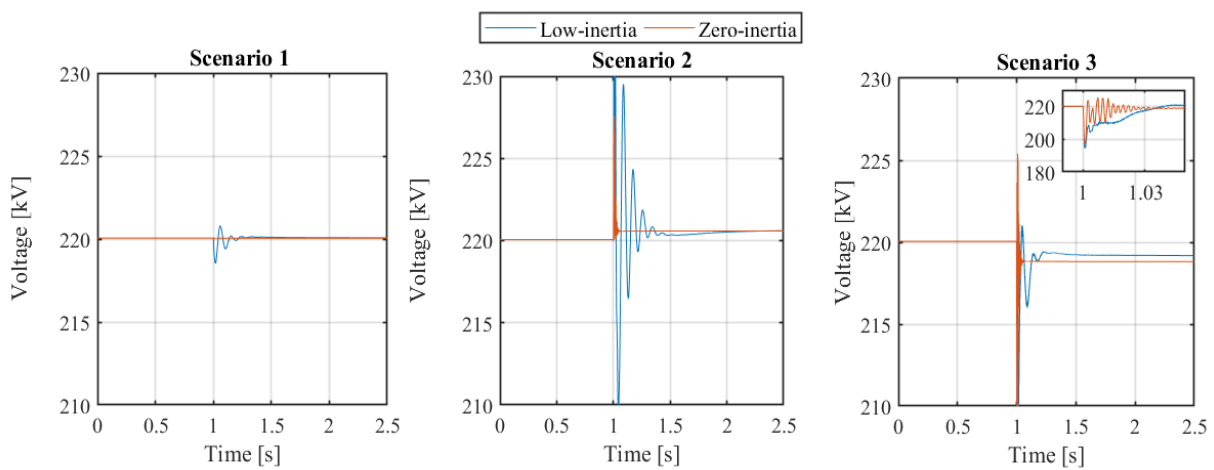

*Figure 6: Voltage deviation at the AC-Hub. Scenario 1: 200 MW power request from a partner TSO. Scenario 2: DC-link outage. Scenario 3: Wind power loss with AC cable disconnection.*

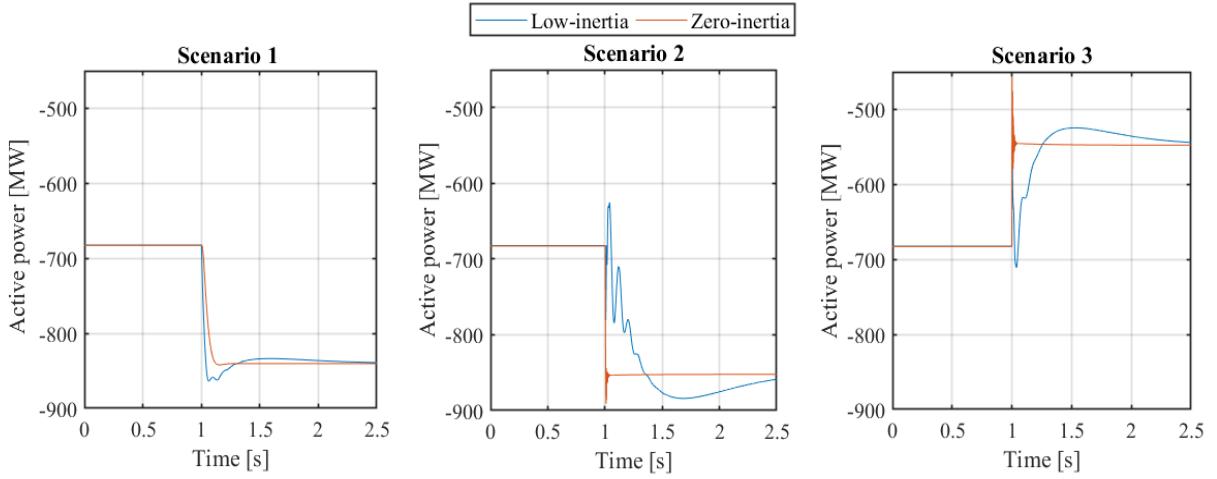

*Figure 7: Active power absorbed by the offshore converters. Scenario 1: 200 MW power request from a partner TSO. Scenario 2: DC-link outage. Scenario 3: Wind power loss with AC cable disconnection.*

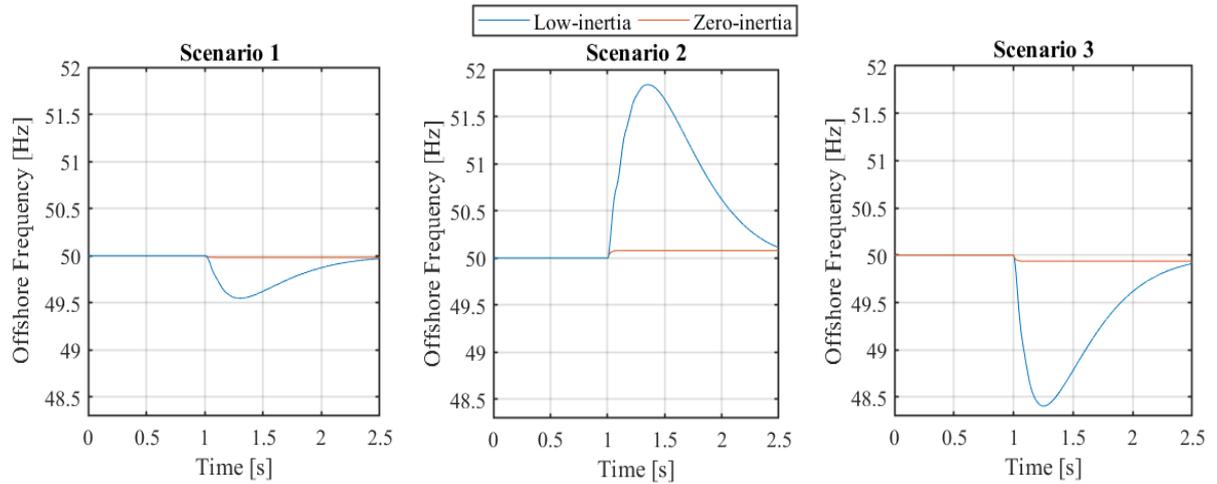

*Figure 8: Offshore frequency deviation. Scenario 1: 200 MW power request from a partner TSO. Scenario 2: DC-link outage. Scenario 3: Wind power loss with AC cable disconnection.*

Regarding the frequency of the offshore system (see Figure 8), in the zero-inertia case the system frequency varies almost instantaneously. However, due to the small values of frequency droops the maximum frequency deviation is small (less than 0.07 Hz). In the low-inertia case, the maximum frequency deviation is higher compared to the zero-inertia case. This is due to the larger values of the frequency droops of the grid-following converters. In the investigated large disturbances, this result in significant frequency deviations (up to 1.6 Hz). Such variations, unacceptable in conventional AC grids, can be tolerated on the isolated offshore island with no load (except for auxiliaries of course). The wind park controllers should be tuned to accept such frequency deviations. Furthermore, the frequency deviation is corrected by the centralized controller, which is updating the power reference set points of the offshore converters.

## V. TOOLS FOR SIMULATION OF A FUTURE NORTH SEA WIND POWER HUB

In this section, we assess the accuracy of the phasor-approximation models for the simulation of offshore systems, such as the NSWPH. The objective is to investigate to what extent those simplified models can be used, and when EMT models are mandatory. The motivation is that system operators of the interconnected AC/DC grids could continue using the phasor approximation for security assessment of their systems including the NSWPH (which is computationally more efficient), as long as the mismatch between the phasor-approximation and the EMT models are within acceptable limits.

As already mentioned, EMT models can capture high frequency modes that phasor-approximation models cannot. Therefore, the appropriateness of the phasor- approximation will depend on the damping ratio of those high frequency modes. The control units that affect the damping ratio of high frequency modes are the virtual impedance controller of the grid-forming converters and the active damping controller of the grid-following converters.

As in Section 4, we are interested in the Hub-voltage and the power exchanged between the offshore converters and the grid. For both configurations, we are interested in evaluating the accuracy of the phasor-approximation for power exchanges between countries, as well as loss of a single element in the NSWPH (N-1 criterion). To this end, we perform the analysis using two scenarios: (i) A 200 MW power request from one of the onshore grids interconnected to the island, and (ii) an offshore HVDC converter outage (N-1 criterion).

Figure 9.a depicts the AC hub voltage (left) and the active power absorbed by an offshore converter (right) following the outage of another offshore converter. During the transients, high frequency oscillations are experienced in both the voltage and the active power. These oscillations are not detected by the phasor-approximation model. However, regarding the active power the mismatch

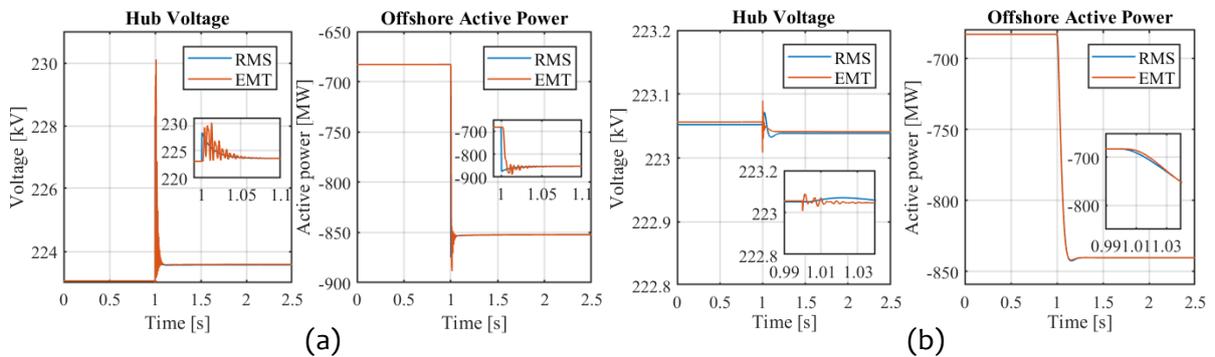

*Figure 9: Accuracy of phasor-approximation model for the zero-inertia configuration: (a) response to an offshore converter outage; (b) response to a power exchange between partner TSOs. Left figures: voltage at the hub. Right figures active power in offshore converter.*

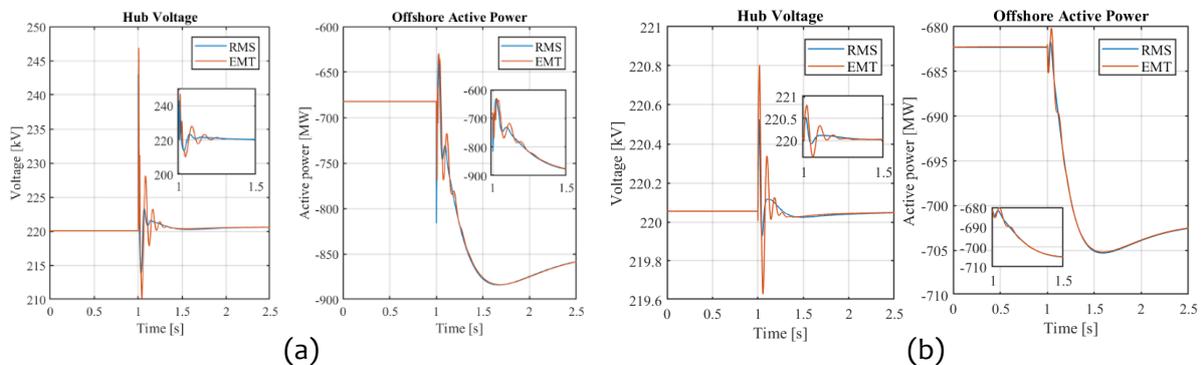

*Figure 10: Accuracy of phasor-approximation model for the low-inertia configuration (a) response to an offshore converter outage; (b) response to a power exchange between partner TSOs. Left figures: voltage at the hub. Right figures: active power in offshore converter*

between the phasor-approximation and the EMT models becomes negligible in less than 5 cycles (0.1 s). Similar conclusions can be drawn for the voltage of the AC hub, although the phasor-approximation model is not able to capture the maximum voltage deviation during the disturbance. Such voltage deviations must be checked with the EMT model.

Similar conclusions can be drawn for the low-inertia case. Figure 10.a depicts the hub voltage (left) and the active power absorbed by an offshore converter (right) following an offshore converter outage. Compared to zero-inertia case, the oscillations are of lower frequency and damp out slower. The mismatch between the phasor-approximation and the EMT models becomes negligible after approximately 10 cycles (0.2 s), which is twice as much compared to the zero-inertia case.

Figure 9.b and Figure 10.b depict the system response during a power exchange between partner TSOs for zero- and low-inertia configurations, respectively. In both cases the mismatch between the phasor-approximation and the EMT models is negligible for the active power absorbed by the offshore converters. As regards the hub voltage a larger mismatch can be observed, due to less-damped eigen-frequencies associated with the cables connected to the AC-Hub bus (these are eigen-frequencies that cannot be controlled by the active damping controllers of the converters). In both cases the phasor- approximation model cannot restitute the reference evolution provided by the EMT model. However, the response of the offshore system, seen from the onshore grids, is accurate enough for evaluating the system response during a power exchange between countries.

Summarizing the results from this section, we can make the following observations:

- Active power disturbances, such as power exchanges between partner TSOs, through the hub, can be accurately simulated using the phasor approximation modelling, in both the zero- and low-inertia configurations.
- As far as the hub voltage is concerned, after large disturbances, the phasor approximation model is much less satisfactory in terms of maximum voltage deviation. Expectedly, it cannot reproduce the oscillations with a period significantly smaller than its time step size. However, the average evolution is rendered satisfactorily.
- During large disturbances, the active power signal is well captured by the phasor-approximation model. Notably, in the zero-inertia case the mismatch between the EMT and the phasor-approximation models becomes negligible in less than five cycles of the alternating current.
- Grid-forming converters can control more tightly the voltage at their point of common coupling, compared to grid-following converters. With proper tuning, they control voltage and active power very well during large disturbances, by providing better damping to high frequency modes, which correspond to state variables associated with the offshore converters. This results in a small mismatch between the phasor-approximation and EMT model, since the dominant dynamics are the ones corresponding to the low frequency modes. The latter are of interest in dynamic security assessment.

## VI. CONCLUSION

To realize the Paris Agreement's target in time, an accelerated deployment of large offshore wind farms is required. To that end, transmission system operators from the North Sea region agreed on exploring the possible development of a North Sea Wind Power Hub (NSWPH). In this paper, first, an economic assessment has been presented, which investigates different system frequencies and voltages for the offshore AC grid. The results show that for a NSWPH the distances between wind farms and the hub are below 50 km. For such short distances, there is no advantage in using low frequency AC systems. Based on a preliminary economic analysis, it is shown that a nominal voltage of 220 kV is preferable for long distances. Only for distances below 20 km, 66 kV is a better option. Beyond this distance, the costs of cables and cable laying dominate the total cost and render the lower voltage option unattractive. These findings suggest that a voltage of 220 kV and a frequency of 50 Hz are the preferred solution for the offshore AC collection grid.

In the technical analysis, this paper explored two configurations for the NSWPH and provided a comparison regarding the dynamic behaviour of such a system. For comparison purposes, performance criteria regarding the voltage and the frequency of the offshore system have been considered, as well as the impact of large disturbances on the active power delivered to the interconnected onshore systems. The following observations were made:

- In the zero-inertia configuration the voltage and the frequency oscillations are better damped for the considered scenarios. Moreover, the active power is transmitted faster between partner TSOs.
- In the zero-inertia configuration, large disturbances, which result in active power imbalance of the offshore system, propagate instantaneously to the interconnected onshore grids and can lead to high frequency deviations in those grids.
- In the low-inertia configuration, such fast propagations of large disturbances are avoided, due to kinetic energy stored in the synchronous condenser. This reduces the impact of offshore incidents on the interconnected onshore grids.

Finally, this paper highlighted that the phasor approximation modelling can be used, as long as eigen-frequencies in power network are well damped. In the zero-inertia case, due to the ability of the grid-forming converter to damp high frequency oscillations, the difference between the phasor approximation and EMT models is negligible. In regard to the low inertia configuration, there is a mismatch between the two models for a slightly higher duration (namely, within 200 ms after the disturbance inception), due to some less-damped eigen-frequencies (depending on the tuning of the inner-current controllers). Our findings suggest that system operators could keep on using the phasor-approximation model in the presence of the NSWPH system for performing dynamic security assessment.

## VII. ACKNOWLEDGMENT

This work was supported by the multiDC project, funded by Innovation Fund Denmark under Grant 6154-00020B.